\documentclass[technote,10pt]{IEEEtran}
\usepackage{cite}
\usepackage{graphicx}
\usepackage{tabularx} 
\usepackage{picinpar}
\usepackage[cmex10]{amsmath}
\usepackage{amsmath,amsfonts,amssymb}
\usepackage{subfigure}
\usepackage{amsthm}
\usepackage{algorithm}
\usepackage{stfloats}
\usepackage{float}
\floatname{algorithm}{{Algorithm}}
\usepackage{bm}
\usepackage{color}
\usepackage{booktabs}
\usepackage{makecell}
\usepackage{mathrsfs}
\usepackage{array}
\usepackage{stfloats}
\usepackage{url}
\usepackage{algpseudocode}
\usepackage{algorithm}
\usepackage{graphicx} 
\usepackage{epstopdf}
\DeclareMathAlphabet{\mathbbold}{U}{bbold}{m}{n}
\usepackage{xcolor} 
\begin{document}
\pdfoptionpdfminorversion=6
\newtheorem{lemma}{Lemma}
\newtheorem{corol}{Corollary}
\newtheorem{theorem}{Theorem}
\newtheorem{proposition}{Proposition}
\newtheorem{definition}{Definition}
\newcommand{\e}{\begin{equation}}
\newcommand{\ee}{\end{equation}}
\newcommand{\eqn}{\begin{eqnarray}}
\newcommand{\eeqn}{\end{eqnarray}}
\title{Deep learning based Channel Estimation and Beamforming in Movable Antenna Systems}

\author{Kaijun Feng, Ziwei Wan, Anwen Liao, Wenyan Ma, Lipeng Zhu, Zhenyu Xiao, Zhen Gao, and Rui Zhang,~\IEEEmembership{Fellow,~IEEE}}

\maketitle

\begin{abstract}
Movable antenna (MA) has emerged as a promising technology for future wireless systems. Compared with traditional fixed-position antennas, MA improves system performance by antenna movement to optimize channel conditions. For multiuser wideband MA systems, this paper proposes deep learning-based framework integrating channel estimation (CE), antenna position optimization, and beamforming, with a clear workflow and enhanced efficiency.  Specifically, to obtain accurate channel state information (CSI), we design a two-stage CE mechanism: first reconstructing the channel matrix from limited measurements via compressive sensing, then introducing a Swin-Transformer-based denoising network to refine CE accuracy for subsequent optimization. Building on this, we address the joint optimization challenge by proposing a Transformer-based network that intelligently maps CSI sequences of candidate positions to optimal MA positions while combining a model-driven weighted minimum mean square error (WMMSE) beamforming approach to achieve better performance. Simulation results demonstrate that the proposed methods achieve superior performance compared with existing counterparts under various conditions. The codes about this work are available at https://github.com/ZiweiWan/Code-4-DL-MA-CE-BF.

\end{abstract}


\section{Introduction}
In the sixth-generation (6G) wireless networks \cite{tns}, massive multiple-input multiple-output (MIMO) transmission has been widely recognized as a key enabling technology \cite{csba,qsr}, while it requires a large number of radio frequency (RF) chains at the base station (BS), leading to substantial hardware costs and power consumption. 
As a remedy, the concept of movable antenna (MA) has recently attracted considerable attention \cite{mapa,6dm}.
	 In practical scenarios, MA improves system performance by physically moving within a predefined spatial range to bypass blockages, adapt to user mobility, and exploit multipath diversity in complex propagation environments, thereby boosting system capacity. Previous studies have demonstrated that compared to traditional MIMO, MA systems require less number of antennas and RF chains to achieve the same or even superior performance, thus alleviating the burden of power consumption and signal processing. Therefore, MA is a promising technique in scenarios like low-altitude communications \cite{uuc} and drone sensing \cite{imm}.
    

To fully harness the potential of MA systems, it is essential to incorporate joint position optimization and beamforming design. The work in \cite{maem} considered MA multiuser transmission with jointly optimal discrete antenna positioning and digital beamforming for sum rate maximization. The authors in \cite{apb} focused on downlink communications for fluid antenna, conceptually similar to MA in terms of flexible antenna positioning, and formulated the MA-associated optimization problem to improve power efficiency under user rate constraints. In \cite{map}, a graph-based MA position optimization framework was introduced. The authors in \cite{mcw} investigated multiuser communication via antenna position optimization with the MAs are employed at the BS. In addition, a probability learning-based cross-entropy optimization (CEO) method was
proposed for optimizing port selection in fluid antenna system\cite{ijt}. However, the works aforementioned were all based on the perfect knowledge of channel state information (CSI), while accurate CSI acquisition is challenging in MA systems. To alleviate this  difficulty, compressive sensing (CS) based channel estimation (CE) schemes have been developed to reconstruct MA/six-dimensional MA channels from a limited number of pilot measurements by exploiting angular-domain sparsity \cite{csb} and user directional sparsity \cite{dce}. However, such methods rely heavily on the accuracy of the assumed sparsity model and may suffer from performance degradation in reality. In addition, most of existing studies on MA-MIMO focus on narrowband communications, leaving a significant research gap on its efficiency in modern wideband orthogonal frequency division multiplexing (OFDM) systems. On the other hand, wireless system design based on deep learning has shown great potential \cite{tca}, which demonstrates strong capabilities by exploiting underlying channel structures. For instance, deep denoising networks such as DnCNN \cite{ddn} and AdaFortiTran \cite{ada} model have been introduced for robust CE in OFDM systems. Recently, the Swin Transformer, a hierarchical vision transformer with shifted-window self-attention \cite{swi}, provides an efficient mechanism for capturing both local and long-range dependencies with manageable complexity, which offers a novel avenue for CE.

In light of this, this paper incorporates deep learning into wideband MA systems.  Unlike existing deep learning based precoding designs with fixed antennas \cite{dlb}\cite{dlbh}, the MA system considered in this paper exploits antenna positioning as a new degree of freedom, which leads to a joint optimization problem involving channel estimation, antenna positioning, and beamforming. The main contributions of this work are summarized as follows:



\begin{itemize}
    \item We propose a two-stage CE framework for accurate CSI acquisition in OFDM-based multiuser MA systems, where CS-based reconstruction from limited measurements is followed by Swin-Transformer-based MA CE network (MA-CENet) denoising to refine the estimation. 
    \item We propose a Transformer-based antenna position optimization method for multiuser systems, which learns a mapping from CSI sequences to optimal MA positions, enabling efficient optimization across subcarriers. 
    \item We propose a model-driven weighted minimum mean square error (WMMSE) beamforming approach for sum-rate maximization in multiuser MA systems, which reduces the iterative complexity of conventional WMMSE while achieving improved performance. 
\end{itemize}


\textit{Notation:} Matrices and column vectors are denoted by upper-case and lower-case bold letters, respectively. Superscripts $(\cdot)^*$, $(\cdot)^{\rm T}$, $(\cdot)^{\rm H}$, $(\cdot)^{-1}$ denote the conjugate, transpose, conjugate transpose and inversion operators, respectively. $\|\mathbf{A}\|_{\rm F}$ is the Frobenius norm of $\mathbf{A}$ and $\|\mathbf{a}\|_p$ is the $l_p$ norm of $\mathbf{a}$. The $(i,j)$-th entry of $\mathbf{A}$ is $[\mathbf{A}]_{i,j}$, and $[\mathbf{A}]_{:,j}([\mathbf{A}]_{i,:})$ denotes the $j$-th row ($i$-th column) of $\mathbf{A}$. The mathematical expectation is denoted by $\mathbb{E}(\cdot)$.

\section{System Model }

\subsection{Channel Model}

We consider a multiuser downlink communication system as shown in Fig. 1. The BS is equipped with $M$ MA elements to serve $K \le M$ users  each equipped with a single fixed antenna. Due to the positioning resolution, we assume that each MA can be located only on $N = N_1 \times N_2$ positions on the $x$-$O$-$y$ plane, where $N_1$ and $N_2$ denote the number of antennas along azimuth and elevation directions, respectively. The coordinate of $n$-th available position is defined as $\mathbf{t}_{n} = [x_{n},y_{n}]^\mathrm T, n = 1\dots N$. Moreover, OFDM with carrier frequency $f_{\rm c}$, bandwidth $B_{\rm s}$, and $N_{\rm c}$ subcarriers with wavelength $\lambda$  is considered. For a wireless channel characterized by $L$ multipath components (MPCs), the $l$-th MPC of the $k$-th user is associated with elevation angle of departure (AoD) $\mathscr{\vartheta}_{l,k}$ , azimuth AoD $ \mathscr{\varphi}_{l,k}$, delay $\tau_{l,k}$, and complex gain $\beta_{l,k},l = 1,2,...,L$ and $k = 1,2,...,K$. For simplicity, we define $\theta_{l,k} \triangleq \sin \vartheta_{l,k} \cos \mathscr{\varphi} _{l,k}
 $ and $ 
\phi_{l,k} \triangleq \cos \vartheta_{l,k}$. The channel $  h_{ k,q}(\mathbf t_n)$ between the $n$-th position and the $k$-th user at the $q$-th subcarrier ($q = 1,2,...,N_{\rm c}$) can be represented as 
\begin{equation}
h_{k,q}(\mathbf t_n) =  \mathbf{d}_k^{\mathrm T}(q)\boldsymbol\Sigma_k   \mathbf g_k(\mathbf {t}_n),
\end{equation}
where 
\[
\begin{aligned}
\mathbf{g}_{k}(\mathbf{t}_{n})
&= [e^{-j\frac{2\pi}{\lambda}\rho_{k,1}(\mathbf{t}_{n})}, \dots,e^{-j\frac{2\pi}{\lambda}\rho_{k,L}(\mathbf{t}_{n})}]^{\mathrm T}\in \mathbb{C}^{ L}, \\
\rho_{k,l}(\mathbf{t}_{n})
&= x_{n}\theta_{k,l} + y_{n}\phi_{k,l},\\
\boldsymbol{\Sigma}_k
&= {\rm diag}\left( \beta_{1,k}, \beta_{2,k},...\beta_{L,k}\right) \in \mathbb{C}^{L \times L}, \\
\mathbf{d}_k(q)&= \left[ e^{-j2\pi \frac{qB_{\rm s} \tau_{1,k}}{N_{\rm c} }}, \dots, e^{-j2\pi \frac{qB_{\rm s} \tau_{L,k}}{N_{\rm c} }} \right]^{\mathrm T} \in \mathbb{C}^{L }.
\end{aligned}
\]
We further define $\bar{\mathbf{h}}[k,q] = $$\left[  h_{ k,q}\left(\mathbf t_{1}\right),   \ldots,  h_{k,q}\left(\mathbf t_{N}\right) \right] ^{\mathrm T}\in \mathbb{C}^{ N}$ as the overall channel vector. 

\section{CE and Denoising}
Building on the system model defined in Section II, we first address the challenge of accurate CSI acquisition which is the essential foundation for subsequent antenna position optimization and beamforming. In this section, we propose the CE scheme for MA by utilizing CS techniques followed by the MA-CENet.

\begin{figure}[t]
\centering
\includegraphics[scale=0.62]{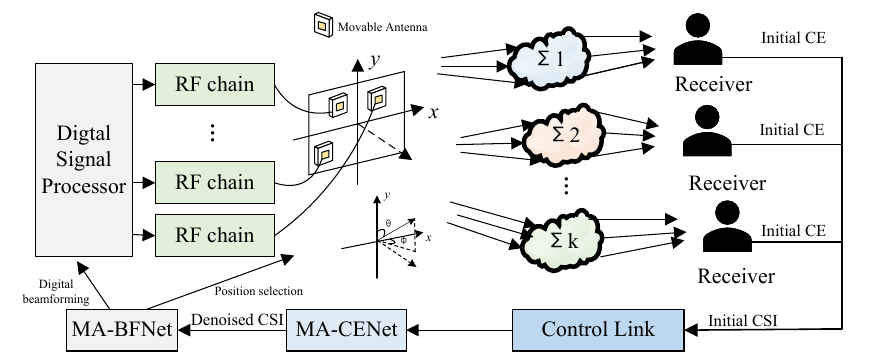}
\caption{MA-based multiuser communication system.}
\label{fig:commSys}
\end{figure}
\subsection{CS-based Initial CE} \label{AA1}
We assume that only one MA is activated at the transmitter side for CE. This is because in absent of CSI, using multiple antennas during CE will not enhance the received signal, and thus will not benefit CE. During CE stage, the MA is moved over $J$ out of $N$ positions with $J \ll N$ and  send pilot signals to all users.  Practically, scheduling all MAs to sequentially visit these $J$  positions reduces the total movement distance, thereby improving time and energy efficiency in practical implementation. We define the set of used positions for CE as ${\cal J} \subseteq \{1,2,...,N\}$ with $\left| {\cal J} \right| = J$, and set ``$1$'' as the known pilot signal without loss of generality. The received signal from the $j$-th used position ($j = 1,2,...,J$) to the $k$-th user at the $q$-th subcarrier is \cite{mapa}
\begin{equation}
y_j[k,q] = \sqrt{P}\,\,
\mathbf{d}_k^{\mathrm T}(q)\boldsymbol\Sigma_k
\mathbf{g}_k(\mathbf{t}_{{\cal J}(j)}) + z_j[k,q],
\end{equation}
where $P$ is the transmit power of the pilot signal, ${\cal J}(j)$ is the $j$-th element of $\cal J$, and $ z _j[k,q]\sim \mathcal{CN}(0,\sigma^2)$ is additive white Gaussian noise (AWGN) with power $\sigma^2$. Stacking all $J$ received signals yields
\begin{equation}
    \hspace{-2mm}\mathbf y[k,q]=[y_1[k,q],\dots,y_J[k,q]]^{\mathrm{T}} = \mathbf {\bf B}^{\rm ce}{\bf A}_k\mathbf x_{k,q} + \mathbf z[k,q], 
    \label{eq:AoD_model}
\end{equation}
where ${\bf{z}}\left[ {k,q} \right] $$=$$ {\left[ {{z_1}\left[ {k,q} \right],{z_2}\left[ {k,q} \right],...,{z_J}\left[ {k,q} \right]} \right]^{\rm{T}}} \in {\mathbb{C}^J},$ $\mathbf{A}_k
$$=$$\big[\boldsymbol{a}(\theta_{1,k},\phi_{1,k}),\dots,\boldsymbol{a}(\theta_{L,k},\phi_{L,k})\big]
\in \mathbb{C}^{N\times L}$ with $\boldsymbol{a}(\theta_l,\phi_l)
= $$\big[e^{-j\frac{2\pi}{\lambda}(x_1\theta_{l,k}+y_1\phi_{l,k})},\dots,e^{-j\frac{2\pi}{\lambda}(x_N\theta_{l,k}+y_N\phi_{l,k})}\big]^{\rm T}
\in \mathbb{C}^N$, ${\bf B}^{\rm ce} = \left[ {\bf I}_N \right]_{{\cal J},:} \in \mathbb{C}^{J \times N}$ is the selection matrix, and  $\mathbf x^{\rm  T}_{k,q} $$\triangleq \sqrt{P}\,\mathbf{d}_k^{\mathrm T}(q)\boldsymbol\Sigma_k$. To resort to state-of-the-art CS techniques, we introduce the dictionary by uniformly discretizing virtual azimuth angle and elevation angle into $G \gg L$ grids to represent the channel in a sparse form. Specifically, the dictionary matrix can be written as
\begin{equation}
   \bar{\mathbf A}=[\boldsymbol{a}(\bar\theta_{1},\bar\phi_{1}),\dots,\boldsymbol{a}(\bar\theta_{g_1},\bar\phi_{g_2}),\dots \boldsymbol{a}(\bar\theta_{G},\bar\phi_{G})\big]\in \mathbb{C}^{N \times G^2},
\end{equation}
where ${\theta _{{g_1}}} =  - 1 + 2{g_1}/G$ and ${\phi _{{g_2}}} =  - 1 + 2{g_2}/G$, $g_1 , g_2 = 1,2,...,G$. Then we can represent the channel as $\mathbf A_k\mathbf x_{k,q} \approx \bar{\mathbf A}\bar{\mathbf x}_{k,q}$ where $\bar{\mathbf x}_{k,q} \in \mathbb{C}^{G^{2}}$ is the sparse coefficient vector. Then, the CE problem at each user can be transformed into the following simultaneous sparse recovery problem: 
\begin{equation}
\begin{aligned}
& \min_{\bar{\mathbf x}_{k,1},\dots,\bar{\mathbf x}_{k,N_{\rm c}}}
 \sum\nolimits_{q=1}^{N_{\rm c}} 
        \left\| {\mathbf y}[k,q] - \mathbf{B}^{\mathrm{ce}}\bar{\mathbf{A}} \bar{\mathbf x}_{k,q} \right\|_2^2
\\
 \text{s.t.} \quad 
    & \| \bar{\mathbf x}_{k,q} \|_0 = L,
\\
& \mathrm{supp}(\bar{\mathbf x}_{k,1}) 
  = \mathrm{supp}(\bar{\mathbf x}_{k,2})
  = \dots
  = \mathrm{supp}(\bar{\mathbf x}_{k,N_{\rm c}}),
\end{aligned}
\label{eq:11}
\end{equation}
and it can be efficiently solved via off-the-shelf CS algorithms. In this paper, we utilize simultaneous orthogonal match pursuit (SOMP) algorithm \cite{otn} to solve problem  \eqref{eq:11}  and obtain the initial CE results as ${\bf \hat h}_{\rm ini}[k,q] \in \mathbb{C}^N$. The details of SOMP are presented in steps 3--12 in Algorithm 1..

\subsection{Swin-Transformer Based Denoising Network} \label{AA
}

After estimating the CSI, the user equipment feeds it back to the BS via the control link for downlink beamforming design, as shown in Fig. 1. To effectively suppress channel noise while preserving the intrinsic spatial structures, we propose the MA-CENet at the BS, as illustrated in Fig. 2. The initial channel estimates $\left\{ {{{{\bf{\hat h}}}_{{\rm{ini}}}}\left[ {k,q} \right]} \right\}_{k = 1,q = 1}^{K,{N_{\rm{c}}}}$ are collected to obtain $\mathcal{H}_\mathrm{est} \in \mathbb{C}^{N\times K\times N_{\rm c}}$. Then, we merge the last two dimensions of $\mathcal{H}_\mathrm{est}$ and separate the real and imaginary parts to form the input vector  $\mathcal{H}_{\mathrm{in}} \in \mathbb{C}^{2\times N\times KN_{\rm c}}$, which forms the input for MA-CENet. As shown in Fig. 2, MA-CENet first passes the input  $\mathcal{H}_{\mathrm{in}} $ through a convolutional layer as patch embedding operation. Next, the embeddings are processed through multiple Swin Transformer Block (STB), which utilizes window-based multi-head self-attention (W-MSA) and shifted window attention to model both local and non-local dependencies. For the $l$-th block, the transformation is formulated as
\begin{equation}
    \mathbf{X}_{l+1} = \mathrm{STB}_l(\mathbf{X}_{l}), 
    \label{eq:stb}
\end{equation}
where each STB consists of a W-MSA module and a feed-forward multilayer perceptron (MLP), both followed by Layer Normalization and residual connections.  To gradually aggregate contextual information, we employ a patch merging strategy that reduces spatial resolution while increasing the channel dimension.
At the decoding stage, we adopt a dual upscaling module to restore the original resolution. Specifically, two parallel branches are used: 1) standard upsampling with convolution, and 2) PixelShuffle-based sub-pixel convolution. 
Finally, the output of MA-CENet $\mathcal{H}_\mathrm{out}\in \mathbb{C}^{2\times N\times KN_{\rm c}}$ is obtained by projecting the features back to the channel. Then, we merge the first dimension as the real-imaginary components of a complex tensor, which is subsequently reshaped to yield the refined channel vectors $\mathcal{H}_\mathrm{den}\in \mathbb{C}^{N\times K\times N_{\rm c}}$. For the proposed denoising network, we choose the normalized mean square error (NMSE) as the loss function, which is defined as
\begin{equation}
\mathrm{NMSE}
= \mathbb{E}\!\left[
\frac{
\sum_{q=1}^{N_{\rm c}}
\left\|
\bigl[ \mathcal{H} \bigr]_{:,:,q}
-
\bigl[ {\mathcal{H}_{\rm den}} \bigr]_{:,:,q}
\right\|_{F}^{2}
}{
\sum_{q=1}^{N_{\rm c}}
\left\|
\bigl[ \mathcal{H} \bigr]_{:,:,q}
\right\|_{F}^{2}
}
\right],
    \label{eq:nmse}
\end{equation}
where $ \mathcal{H}\in \mathbb{C}^{N\times K\times N_{\rm c}}$ denotes the perfect CSI.

\begin{algorithm}[t]

\caption{ {Deep learning based CE and beamforming framework for multiuser wideband MA systems}}
\label{alg:DL_MA_framework}
\begin{algorithmic}[1]

\Statex \vspace{0.15em}
\Statex \textbf{Part I: Initial CE}
\State \textbf{Input:} Pilot observations $\{\mathbf{y}[k,q]\}_{k = 1,q = 1}^{K,{N_{\rm{c}}}}$, $\forall k,q$; Selection matrix $\mathbf{B}^{\mathrm{ce}}$; Dictionary $\bar{\mathbf{A}}$; 
\For{$k=1$ to $K$}
    \State \textbf{SOMP for $\hat{\mathbf A}_k$ reconstruction}
    \State Form sensing matrix $\boldsymbol{\Phi}\triangleq \mathbf B^{\rm ce}\bar{\mathbf A}\in\mathbb C^{J\times G^2}$.
    \State Stack observations $\mathbf Y_k\triangleq [\mathbf y[k,1],\ldots,\mathbf y[k,N_c]]\in\mathbb C^{J\times N_c}$.
    \State Initialize support set $\mathcal S\leftarrow\emptyset$, residual matrix $\mathbf R\leftarrow \mathbf Y_k$.
    \For{$t=1$ to $L$}
        \State $g^\star \leftarrow \arg\max\limits_{g\notin\mathcal S}\left\| \boldsymbol{\phi}_g^{\mathrm H}\mathbf R\right\|_2$, where $\boldsymbol{\phi}_g$ is the $g$-th column of $\boldsymbol{\Phi}$.
        \State $\mathcal S \leftarrow \mathcal S \cup \{g^\star\}$.
        \State $\mathbf X_{\mathcal S}\leftarrow \boldsymbol{\Phi}_{\mathcal S}^{\dagger}\mathbf Y_k
        =(\boldsymbol{\Phi}_{\mathcal S}^{\mathrm H}\boldsymbol{\Phi}_{\mathcal S})^{-1}\boldsymbol{\Phi}_{\mathcal S}^{\mathrm H}\mathbf Y_k$.
        \State $\mathbf R \leftarrow \mathbf Y_k-\boldsymbol{\Phi}_{\mathcal S}\mathbf X_{\mathcal S}$.
    \EndFor
    \State Map each $g\in\mathcal S$ to grid pair $(g_1,g_2)$ and set $\hat\theta_{l,k}\!=\!\bar\theta_{g_1}$, $\hat\phi_{l,k}\!=\!\bar\phi_{g_2}$, where $\bar\theta_{g_1}=-1+\frac{2g_1}{G}$, $\bar\phi_{g_2}=-1+\frac{2g_2}{G}$.
    \State Reconstruct $\hat{\mathbf A}_k \leftarrow [\mathbf a(\hat\theta_{1,k},\hat\phi_{1,k}),\ldots,\mathbf a(\hat\theta_{L,k},\hat\phi_{L,k})]$.
    \For{$q=1$ to $N_c$}
        \State Applying least square (LS) algorithm  $\hat{\mathbf x}_{k,q}\leftarrow \arg\min_{\mathbf x}\|\mathbf y[k,q]-\mathbf B^{\rm ce}\hat{\mathbf A}_k\mathbf x\|_2^2$.
        \State $\hat{\mathbf x}_{k,q}\leftarrow(\mathbf B^{\rm ce}\hat{\mathbf A}_k)^{-1}\mathbf y[k,q]$.
        \State  $\hat{\mathbf h}_{\mathrm{ini}}[k,q]\leftarrow \hat{\mathbf A}_k\,\hat{\mathbf x}_{k,q}\in\mathbb C^{N}$.
    \EndFor
\EndFor
\State \textbf{Output:} The estimate CSI $\left\{ {{{{\bf{\hat h}}}_{{\rm{ini}}}}\left[ {k,q} \right]} \right\}_{k = 1,q = 1}^{K,{N_{\rm{c}}}}$

\Statex \vspace{0.15em}
\Statex \textbf{Part II: CSI denoising}
\State \textbf{Input:} The estimate CSI $\left\{ {{{{\bf{\hat h}}}_{{\rm{ini}}}}\left[ {k,q} \right]} \right\}_{k = 1,q = 1}^{K,{N_{\rm{c}}}}$
\State Stack $\{{\mathbf{\hat h}}_{\mathrm{ini}}[k,q]\}$ as $\mathcal{H}_{\mathrm{est}}\in\mathbb{C}^{N\times K\times N_c}$ at the BS.
\State Form $\mathcal{H}_{\mathrm{in}}$ (split real/imag and reshape).
\State \textbf{Output:} $\mathcal{H}_{\mathrm{den}}\leftarrow \mathrm{MA\text{-}CENet}(\mathcal{H}_{\mathrm{in}})$.

\Statex \vspace{0.15em}
\Statex \textbf{Part III: Position selection \& beamforming}
\State \textbf{Input:} The denoised CSI $\mathcal{H}_{\mathrm{den}}$.
\State $\tilde{\mathbf{B}}\leftarrow \mathrm{MA\text{-}PSN}(\mathcal{H}_{\mathrm{den}})$.
\State $\mathbf{B}^{\rm pos}\leftarrow \mathrm{Argmax}\!\left(\mathrm{Softmax}(\tilde{\mathbf{B}})\right)\in\{0,1\}^{N\times M}$.
\State $\mathbf{h}^{\rm T}_{\mathrm{equ}}[k,q]\leftarrow \mathbf{h}^{\rm T}_{\mathrm{den}}[k,q]\mathbf{B}^{\rm pos}$, $\forall k,q$; stack to $\mathcal{H}_{\mathrm{equ}}$.
\State \textbf{Output:} Predict beamforming matrix $\{\mathbf{W}[q]\}_{q=1}^{N_c}\leftarrow \mathrm{MA\text{-}DBN}(\mathcal{H}_{\mathrm{equ}})$.
\end{algorithmic}
\end{algorithm}

\section{MA Position And Beamforming Optimization}
\begin{figure*}[t]
\centering
\includegraphics[scale=0.33]{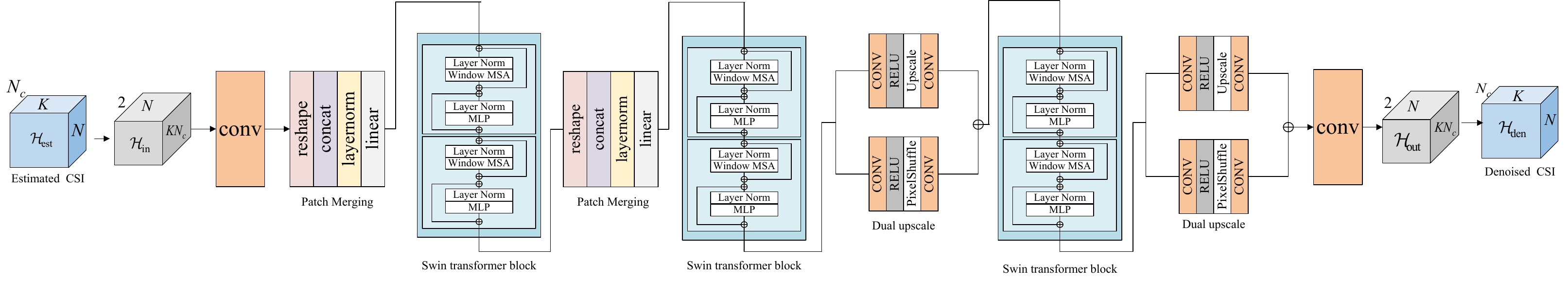}
\caption{ The proposed MA-CENet conducts denoising to the initial estimate of channel to improve estimation performance.}
\label{fig:neuMod}

\end{figure*}
The denoised CSI $\mathcal{H}_\mathrm{den}$ given in Section III serves as the input for MA position optimization detailed in this section. The overall procedure of the proposed framework is summarized in steps 26--30 in Algorithm 1.  Accurate CSI characterizes the channel gains between candidate positions and all users, enabling accurate performance evaluation for selected antenna positions.

In this section, we introduce the Transformer-based MA beamforming network (MA-BFNet). The overall workflow of the MA-BFNet is illustrated in Fig. 3, which consists of MA position selection network (MA-PSN) and MA digital beamforming network (MA-DBN). At this stage, all $M$ MAs at the BS are used and we define a position-selection matrix $\mathbf{B}^{\mathrm{pos}} \in \{0,1\}^{N\times M}$ which is shared by all the subcarriers.  Each column (row) of $\mathbf{B}^{\mathrm{pos}}$ contains only one non-zero element, which indicates the index of the selected positions. The equivalent channel after position selection ${\mathbf{h}^{\mathrm T}}[k,q]\in \mathbb{C}^{M}$ can be expressed as 
\begin{equation}
\mathbf {h}^{\mathrm T}[k,q] = \bar{\mathbf{h}}^{\mathrm T}[k,q]\mathbf{B}^{\mathrm{pos}}.
\end{equation}
Therefore, the received signal of the $k$-th user at the $q$-th subcarrier is given by
\begin{equation}
 y[k,q] = {\mathbf{h}}^{\mathrm T}[k,q]\mathbf {W}[q]\mathbf{s}[q] + z[k,q],
\end{equation}
where $\mathbf{W}[q] \in \mathbb{C}^{M\times K}$ is the beamforming matrix with power constraint $\|\mathbf W[q]\|_{\rm F}^2 = P_{\rm t}, \ \forall q$, $P_{\rm t}$ is the transmit power, $\mathbf{s}[q] \in \mathbb{C}^{K}$ is the transmitted signal satisfying $\mathbb{E}\!\bigl(\mathbf s^{\mathrm H}[q]\,\mathbf s[q]\bigr)=\mathbf {I}_{K}$, and $z[k,q] \sim \mathcal{CN}(0, \sigma^2)$  is AWGN. The sum rate $R$ is adopted as the objective function for beamforming design, which can be expressed as 
\begin{equation}
R = \frac{1}{N_{\rm c}} 
\sum\nolimits_{k=1}^{K}
\sum\nolimits_{q=1}^{N_{\rm c}}
\log_{2}\!\big(1 + \mathrm{SINR}[k,q]\big).
\end{equation}
In (10), the signal-to-interference-plus-noise ratio (SINR) is
\begin{equation}
\mathrm{SINR}[k,q] =
\frac{\left|\mathbf{h}^{\rm T}[k,q]\mathbf{B}^{\mathrm{pos}}\mathbf w[k,q]\right|^{2}}
{\displaystyle \sum\nolimits_{i=1,i\neq k}^{K}
\left|\mathbf{h}^{\rm T}[k,q]\mathbf{B}^{\mathrm{pos}} \mathbf w[i,q]\right|^{2}
+ \sigma^{2}},
\end{equation}
with $\mathbf w[k,q]$ denotes the $k$-th column of $\mathbf{W}[q]$.

\subsection{MA-PSN Design}
As illustrated in Fig. 3, the proposed MA-PSN employs a self-attention mechanism to extract cross-subcarrier dependencies from the input CSI $\mathcal{H}_\mathrm{den}$, enabling intelligent selection of optimal MA positions in multiuser systems to enhance overall sum rate. The design of $\mathbf{B}^{\mathrm{pos}}$ can be formulated as an $N$-class classification task, where each of the $M$ antennas is assigned to the most suitable position. First, the input denoised CSI $\mathcal{H}_\mathrm{den}$ is converted into a real-valued vector by concatenating its real part and imaginary part, and then reshaped into a matrix $\mathbf {H} \in \mathbb{R}^{N_{\rm c} \times 2K N}$. Then, $\mathbf H$ is used as the input to Transformer Encoder 1. The extracted features with dimensions  are passed through the fully connected (FC) linear layer and reshaped to obtain $\tilde{\mathbf{B}} \in \mathbb{R}^{N \times M}$, and each element in $\tilde{\mathbf{B}}$ indicates the probability of an antenna being assigned to a specific location. Then, the softmax activation function is applied to convert $\tilde{\mathbf{B}}$ into a probability matrix. To obtain the position optimization matrix, we use the argmax function to determine the maximum value in each column of matrix $\mathbf{B}^{\mathrm{pos}}$. More specifically, for the $[i, j]$-th entry  of $\mathbf{B}^{\mathrm{pos}}$, it equals 1 only when  ${i} = \arg\max(\text{softmax}([\tilde{\mathbf{B}}]_{:,j}))$ and equals 0 otherwise. Here, we implement a sequential selection mechanism to enforce the “unique position" constraint: for each antenna (column) in turn, we select the position with the highest score, then set the scores of this selected row to zero across all remaining columns.  This design prevents the issues such as antenna overlap or duplicate assignments. Through this method, we ensure that each antenna can select its most suitable position, thereby maximizing the system sum rate.

\subsection{MA-DBN Design}
\begin{figure}[t]
\centering
\includegraphics[scale=0.44]{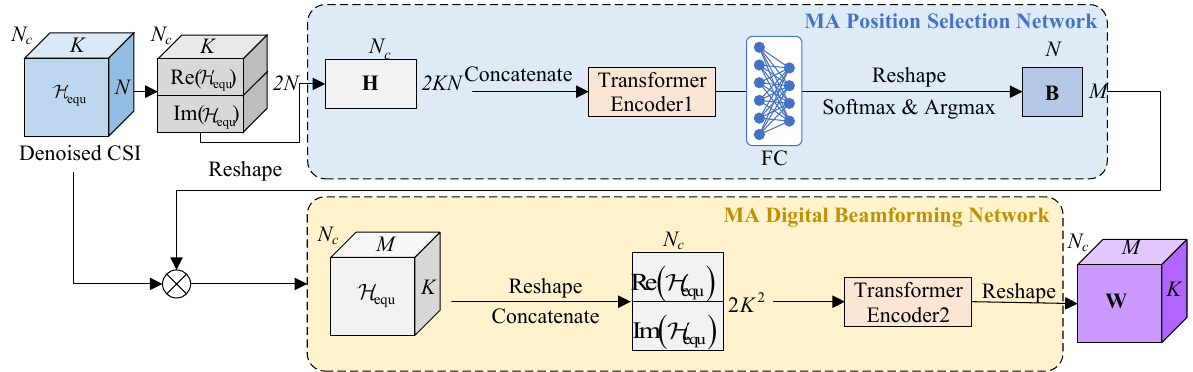}
\caption{The proposed MA-BFNet integrating antenna position optimization and beamforming. }
\label{fig:ma_system}
\end{figure}
As illustrated in Fig. 3, the proposed MA-DBN integrates an encoder with the WMMSE algorithm, learning equivalent channel features to adaptively design the beamforming matrix in a model-driven way. Given the  optimized position matrix $\mathbf{B}^{\mathrm{pos}}$ and the denoised CSI $\mathcal{H}_\mathrm{den}$, we split $\mathcal{H}_\mathrm{den}$ along  the dimensions of the users and the subcarriers into $N_{\rm c}\times K$ slices $\left\{ {{{{\mathbf h}_{\rm den}}}\left[ {k,q} \right]} \right\}_{k = 1,q = 1}^{K,{N_{\rm{c}}}}$, and further define the equivalent baseband channel matrix as $\mathbf h_{\text{equ}}^\mathrm T[k,q] =\mathbf{h}_{\rm den}^\mathrm T[k,q] \mathbf{B}^{\mathrm{pos}}$. The optimization problem for MA digital beamforming can be formulated as

\begin{equation}
\begin{aligned}
\label{equ:Rcs}
    \max_{{\mathbf W}[q]} \quad & R \\
    \text{s.t.} \quad & \|\mathbf W[q]\|_{\rm F}^2 = P_{\rm t}, \ \forall q.
\end{aligned}
\end{equation}
In the traditional WMMSE algorithm, problem \eqref{equ:Rcs} can be further decomposed into three convex subproblems, which are solved alternately to obtain a locally optimal digital beamforming matrix. Specifically, the alternate optimization of WMMSE can be summarized as follows:

\begin{align}
\hspace{-4mm}u[k, q] &= 
\frac{\mathbf h_{\text{equ}}^\mathrm T[k, q]\mathbf w[k, q]}
{\sum_{p=1, p \neq k}^{K} |\mathbf {h}^\mathrm T_{\text{equ}}[k, q]\mathbf  w[p, q]|^2 + \sigma^2}, \\[4pt]
\hspace{-4mm}v[k,q] & = 
\left( 1 - 
\frac{|\mathbf h^\mathrm T_{\text{equ}}[k,q]\mathbf w[k,q]|^2}
{\sum_{p=1, p \neq k}^K 
|\mathbf h^\mathrm T_{\text{equ}}[k,q] \mathbf w[p,q]|^2 + \sigma^2} 
\right)^{-1}, \\[4pt]
\hspace{-4mm}\mathbf w[k,q] &=
\Big( 
\mu[q]\mathbf I_K + 
\sum\nolimits_{p=1}^{K} v[p,q] |u[p,q]|^2 
\mathbf h_{\text{equ}}[p,q] \nonumber \\
&\quad \times \mathbf h_{\text{equ}}^\mathrm H[p,q]
\Big)^{-1} 
u[k,q] v[k,q] \mathbf h_{\text{equ}}[k,q],
\end{align}
where $u[k,q]$ and $v[k,q]$ are the corresponding auxiliary factors, and $ \mu[q]$ is the Lagrange multiplier. Traditional WMMSE can theoretically converge to a good solution but needs many iterations with slow convergence   \cite{GZ}. To overcome this, we introduce a Transformer-based model-driven WMMSE algorithm for designing $\mathbf w[k,q]$. According to (14), the optimal value of $\mathbf w[k,q]$ takes the following form:
\begin{align}
\mathbf w[k, q] = & \left( b[q] \mathbf I_K + \sum\nolimits_{p=1}^{K} c[p, q] \mathbf h_{\text{equ}}[p,q] \mathbf h_{\text{equ}}^\mathrm H[p, q]\right)^{-1} \nonumber \\
    & \times a[k, q] \mathbf h_{\text{equ}}[k, q],
\end{align}
which indicates that the optimal digital beamforming matrix depends on parameters $\{ a[k,q], b[q], c[k, q], \}$, $\forall k, \forall q$. To obtain their values without a large number of iterations, we propose a model-driven network. As shown in Fig. 3, the MA-DBN of MA-BFNet takes the equivalent channels $ \mathcal H_{\text{equ}} \in \mathbb{C}^{N_{\rm c} \times K \times M}$ which is formed by $\left\{ {{{{\mathbf{ h}_{\rm equ}}}}\left[ {k,q} \right]} \right\}_{k = 1,q = 1}^{K,{N_{\rm{c}}}}$ as the input, then, by reshaping and concatenating the real and imaginary parts of $\mathcal H_{\text{equ}}$, we convert $\mathcal H_{\text{equ}}$ into a real-valued input sequence. After that,  the sequence obtained above is processed by Transformer encoder 2 to output the beamforming matrix $\mathbf W[q],\quad q = 1, 2,\dots, N_{\rm c}$. For the proposed MA-BFNet, we choose the negative sum rate as the loss function, given by $L_{\text{loss}} = -R$.

\section{Simulation Results}
\begin{figure}[t]
\centering
\includegraphics[scale=0.35]{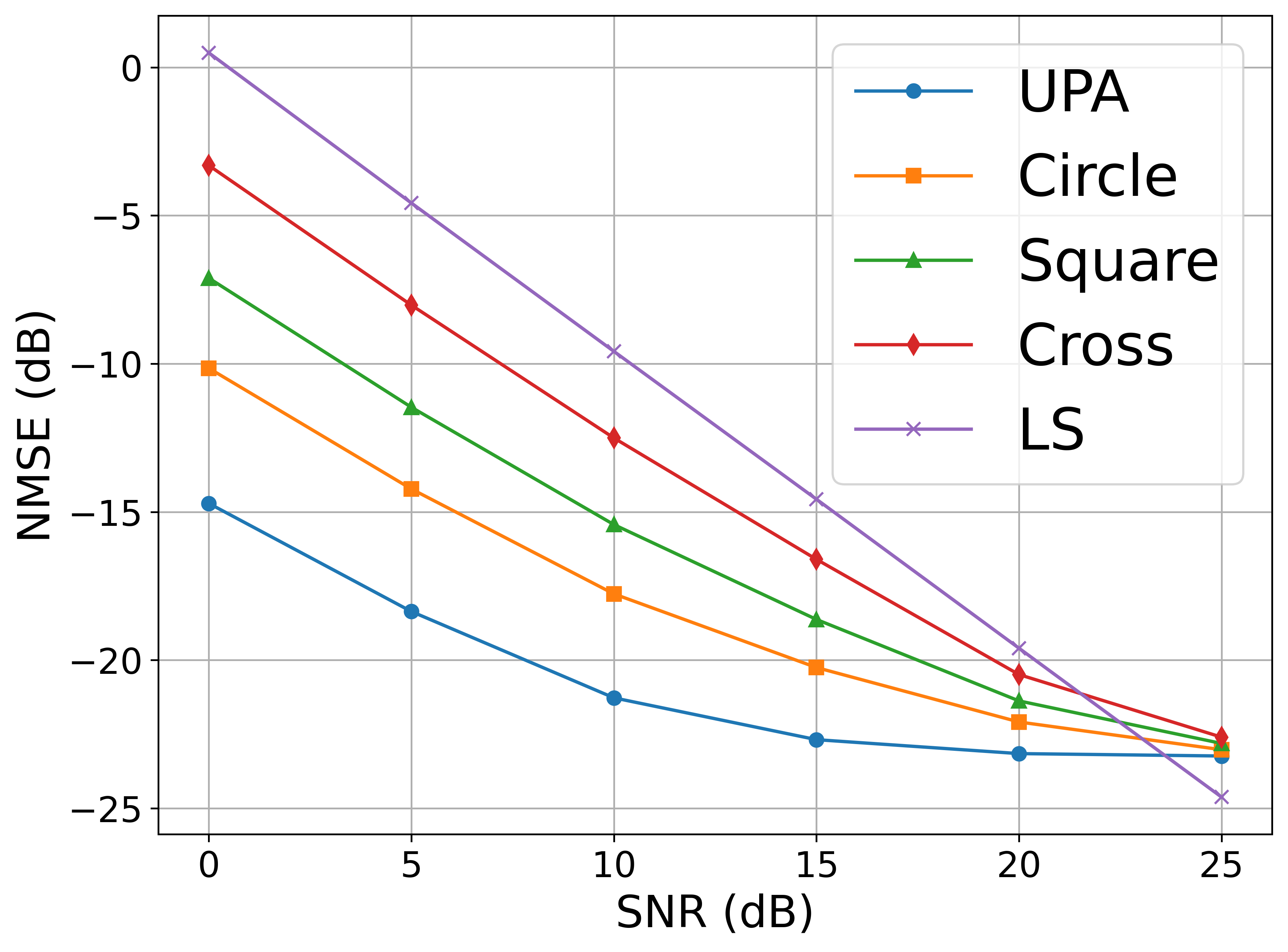}

{%
\makeatletter
\renewcommand{\fnum@figure}{{Fig.~\thefigure}}
\makeatother
\caption{{The performance comparison of NMSE versus SNR for various MA positioning setups.}}
}

\label{fig:commSys}
\end{figure}
\begin{figure}[t]
    \centering
    \subfigure[]{%
        \includegraphics[width=0.49\linewidth]{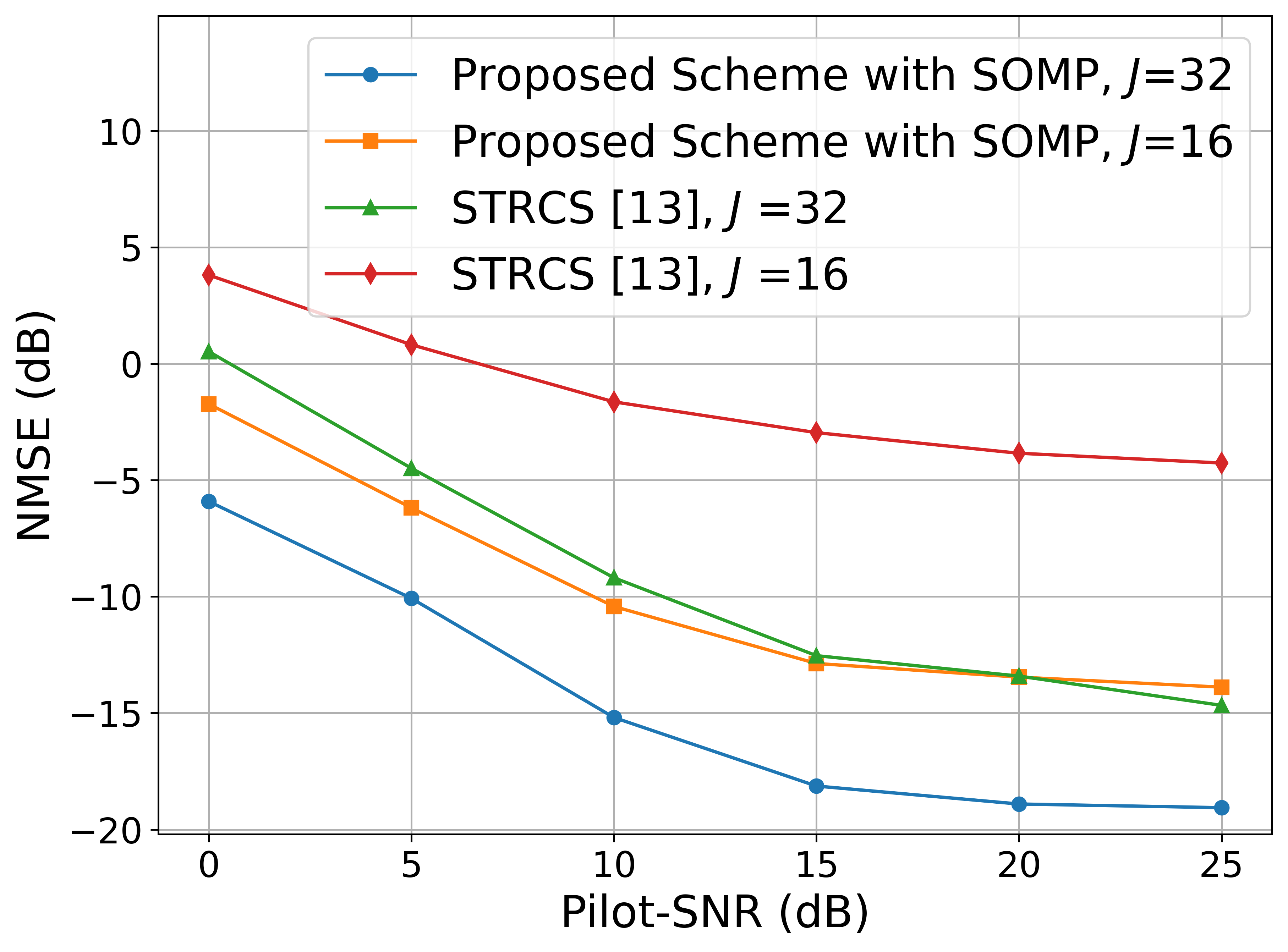}%
        \label{fig:nmse_pos}%
    }\hfill
    \subfigure[]{%
        \includegraphics[width=0.49\linewidth]{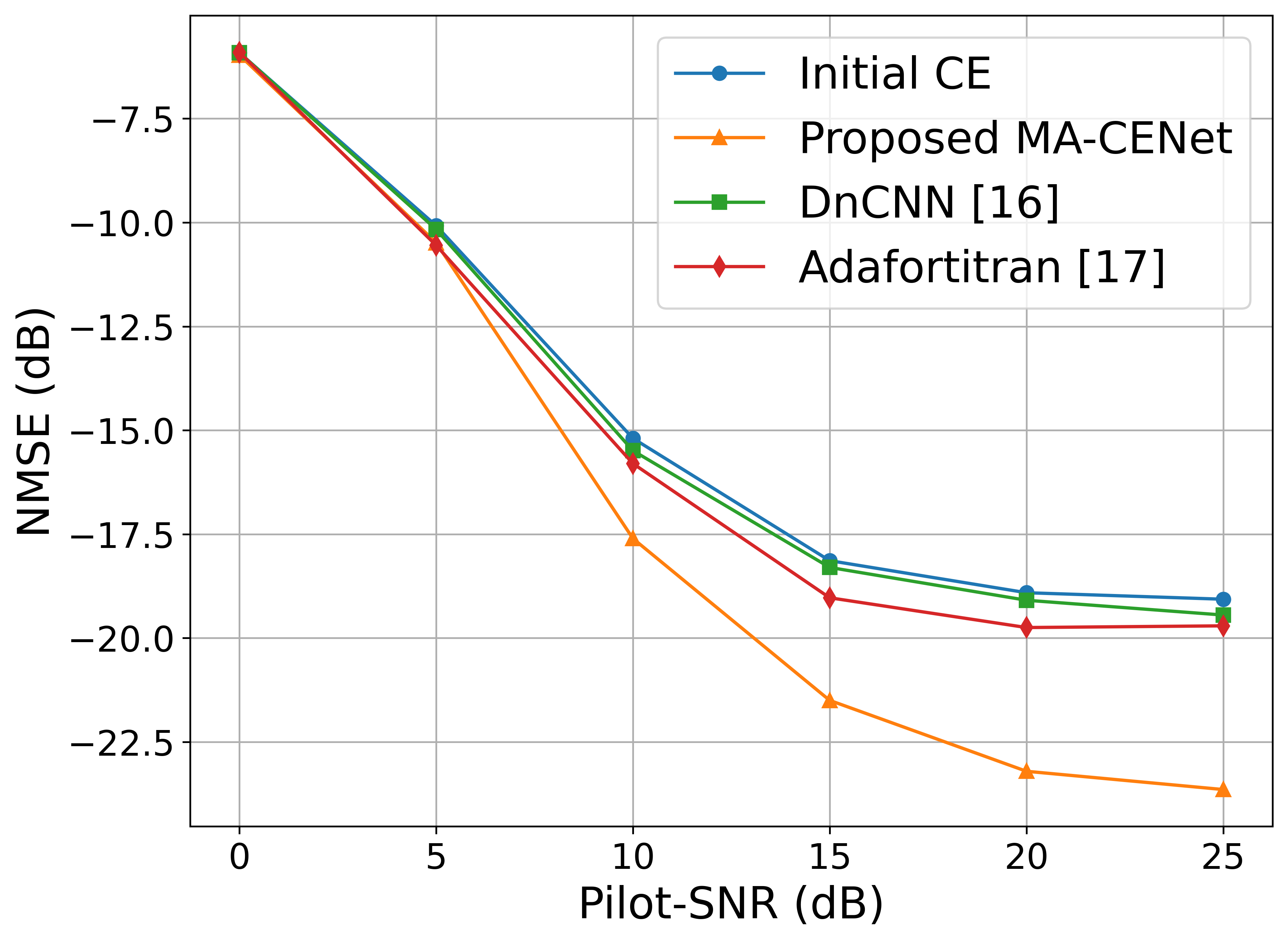}%
        \label{fig:nmse_dl}%
    }
    \caption{NMSE performance for CE: 
    (a) NMSE versus Pilot-SNR under different CE algorithms; 
    (b) NMSE versus Pilot-SNR under different deep learning-based denoising networks.}
    \label{fig:combined_nmse}
\end{figure}

\begin{figure*}[t]
  \centering
  \subfigure[]{\includegraphics[scale=0.176]{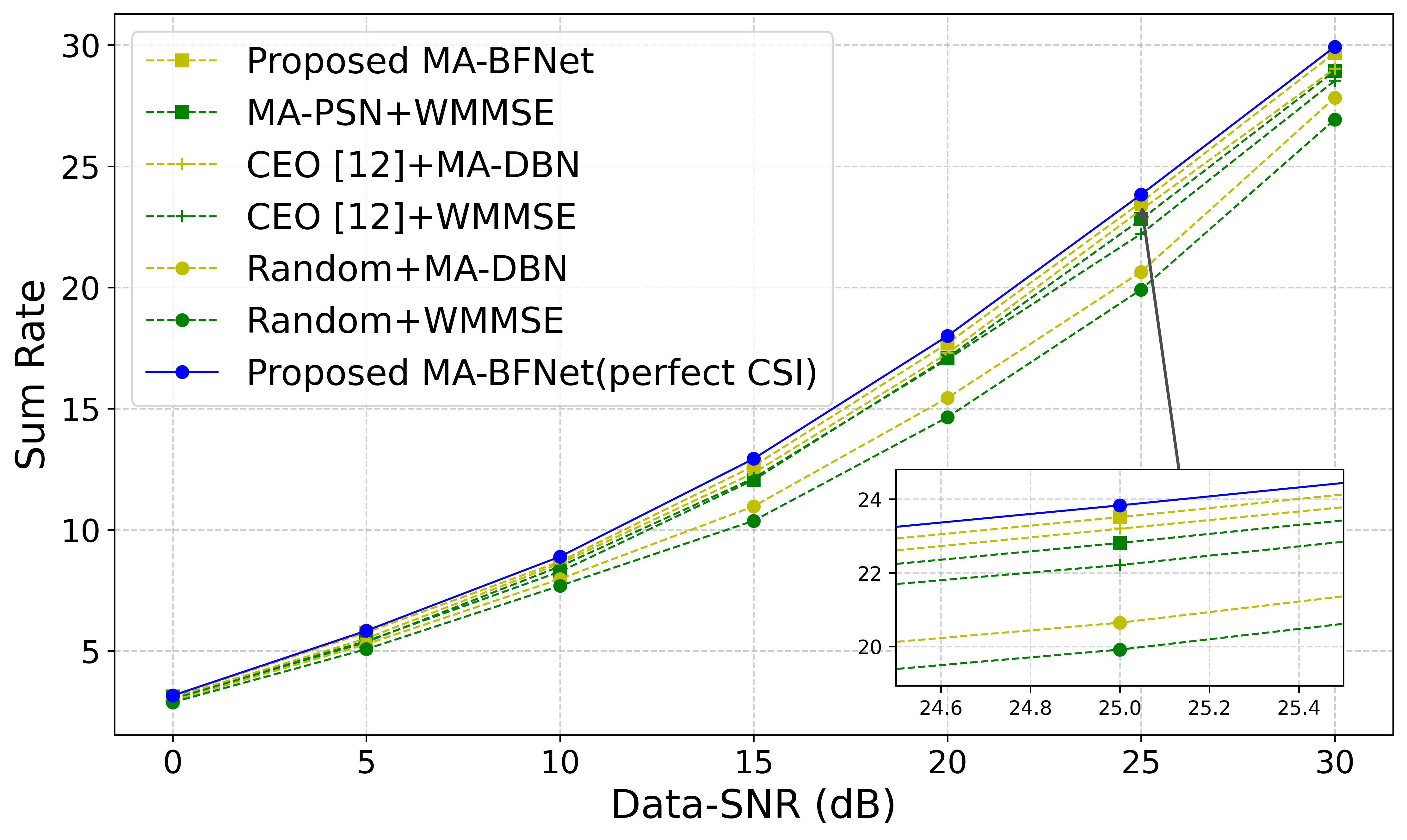}\label{fig2:nmse}}%
  \subfigure[]{\includegraphics[scale=0.176]{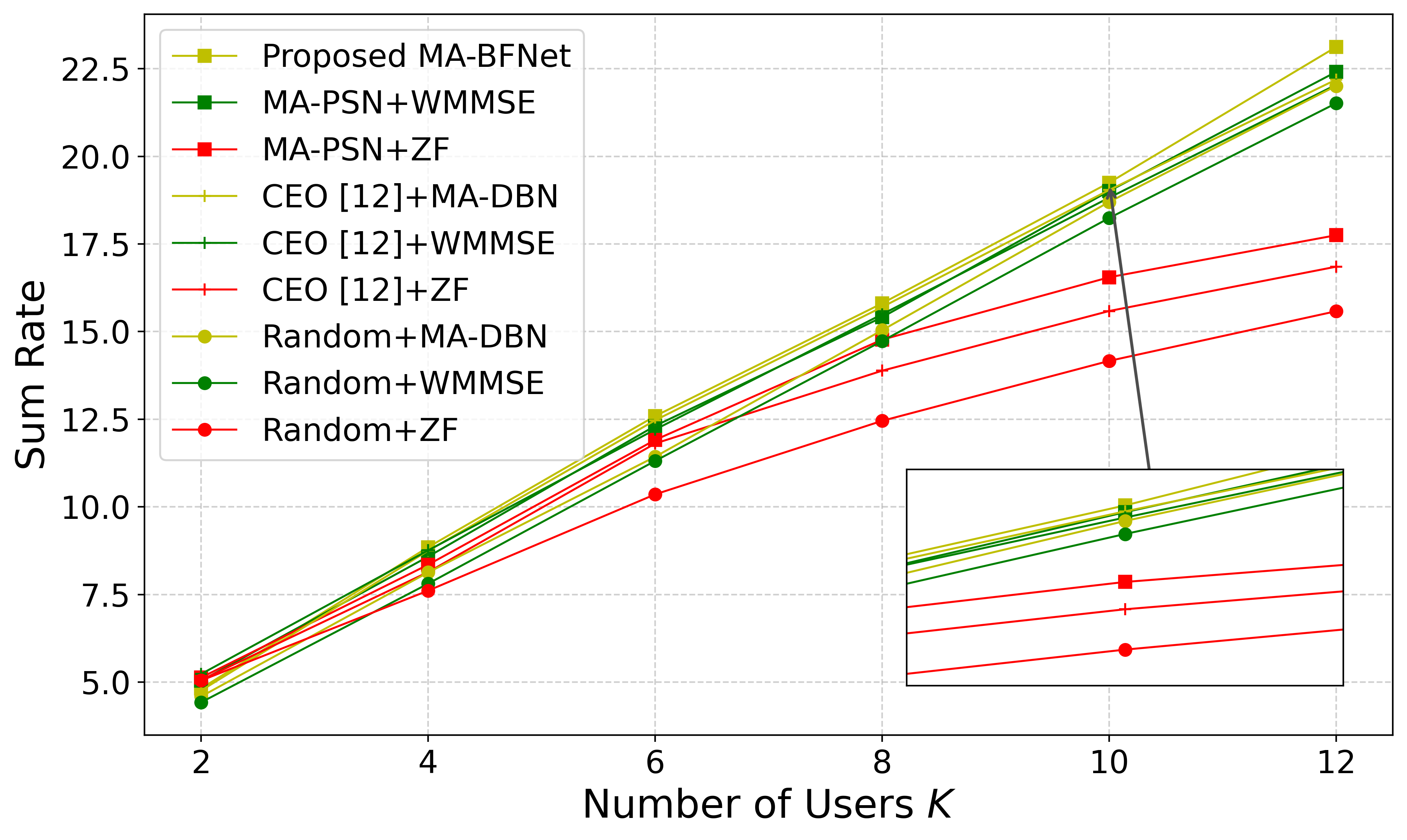}\label{NMSEvsuser}}
   \subfigure[]{\includegraphics[scale=0.176]{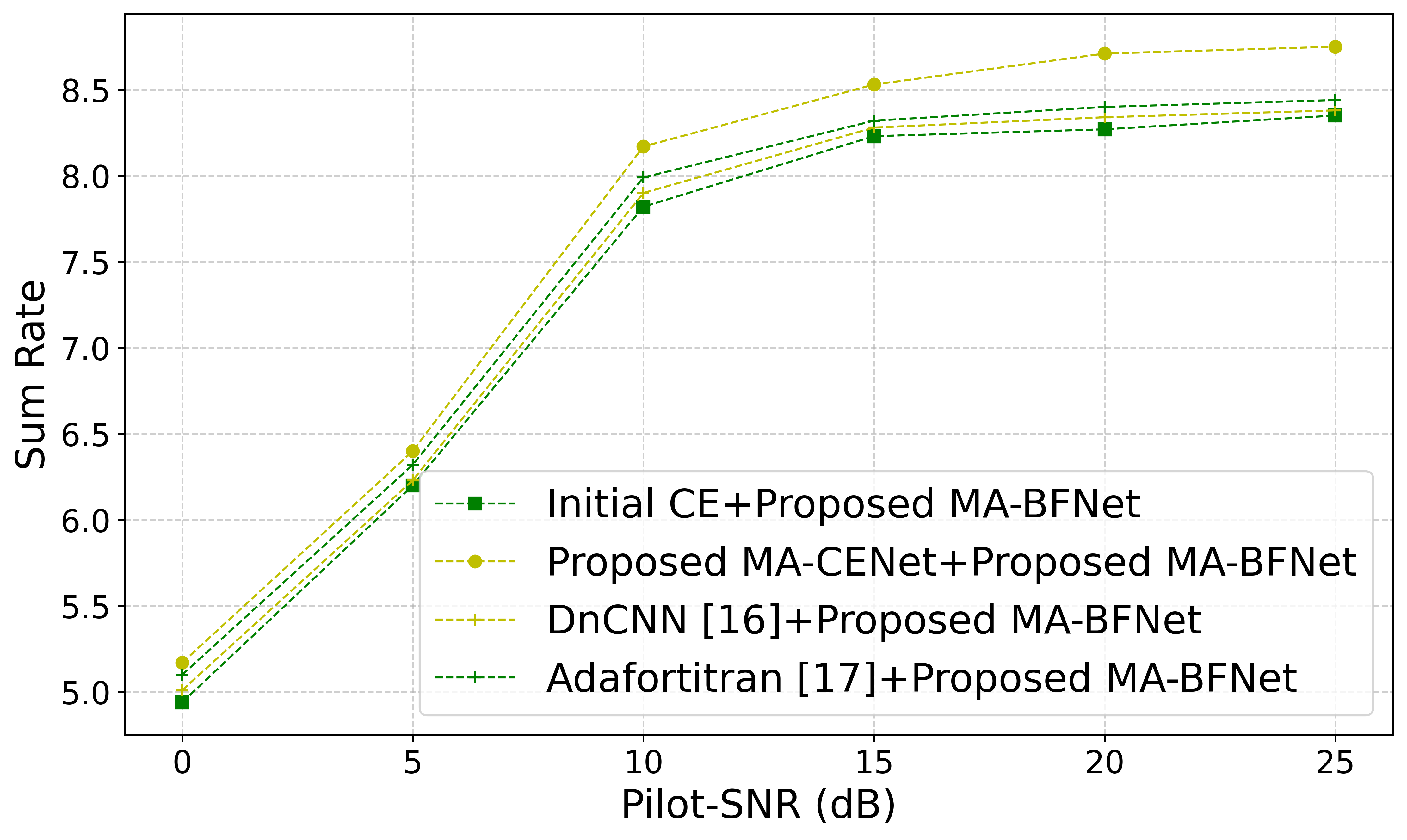}}\label{NMSEvsce}
    \subfigure[]{\includegraphics[scale=0.176]{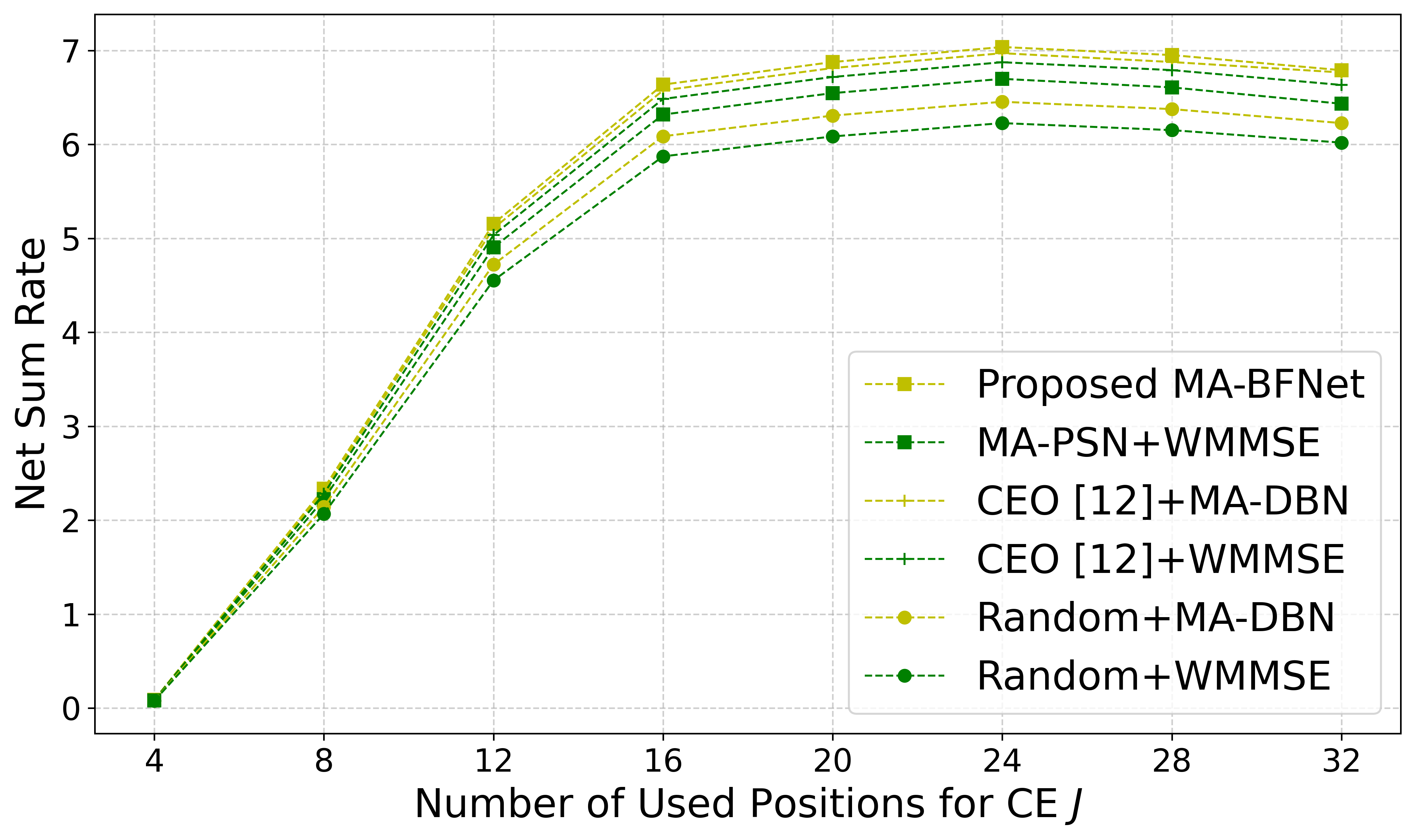}\label{NMSEvspil}}
  \caption{Performance evaluation for MA system:  (a) Sum rate versus Data-SNR; (b) Sum rate versus the number of users; (c) Sum rate versus Pilot-SNR; (d) Net sum rate versus the number of pilots. \label{fig:ma_all}}
\end{figure*}

In simulations, unless stated otherwise, we set $f_{\rm c} = 30$\,GHz, $B_{\rm s} = 30$\,MHz, $N = 8 \times8$, $M = 4$, $K = 4$, $J = 32$ and $N_{\rm c} = 32$. For channel model, we assume $L = 6$ and the AoDs are independent and identically distributed within $(-\tfrac{\pi}{2},\tfrac{\pi}{2}\bigr)$. Moreover, $\beta_{l,k} \sim \mathcal{CN}\!\left(0, \frac{1}{L}\right)$, and $\tau_{l,k}$'s are uniformly distributed within $[{0, \tfrac{8}{B_{\rm s}}}]$. MA-CENet and MA-BFNet are trained using the Adam optimizer with a batch size of 512 for 100 epochs, along with a learning-rate warm-up strategy. We define pilot-SNR as $P/\sigma^2$ and data-SNR as $P_{\rm t}/\sigma^2$.


 Fig. 4 illustrates the NMSE performance of the initial CE under different MA movement patterns. Specifically, four MA position setups are considered. The movement settings are the same as \cite{csb}. It is observed that the uniform planar array (UPA)-shaped setup achieves the best channel estimation accuracy. Accordingly, the initial CSI obtained under the UPA-shaped movement is adopted in subsequent simulations.
Fig.~5(a) compares the NMSE versus Pilot-SNR for CS-based initial CE. It can be observed that the proposed scheme incorporating multiple subcarriers yields a noticeable reduction of NMSE compared to the STRCS algorithm \cite{csb}.
Fig.~5(b) focuses on the impact of the subsequent denoising network and plots the NMSE versus Pilot-SNR for different denoising network. We select DnCNN \cite{ddn} and AdaFortiTran \cite{ada} as two denoising approaches for comparative analysis. We can observe that the proposed MA-CENet consistently achieves the lowest NMSE, which indicates it can more effectively suppress noise and recover finer channel structures than DnCNN and the AdaFortiTran. For example, MA-CENet exhibits about $3$\,dB NMSE improvement over other schemes in the practical Pilot-SNR regime from $0-30$ dB. which facilitates the subsequent beamforming design.

For beamforming performance evaluation, we choose the estimated channel obtained at Pilot-SNR $=10$ dB as the input of the proposed MA-BFNet. For comparison with MA-PSN, we adopt random selection and CEO algorithm \cite{ijt}. For comparison with MA-DBN, we adopt the traditional WMMSE algorithm \cite{aiw} and zero forcing (ZF) algorithm. 
Fig.~6(a) shows the sum rate of the considered schemes versus Data-SNR. It can be observed that the proposed MA-BFNet achieves the best performance across the whole Data-SNR range, and the ablation studies confirm the effectiveness of the proposed MA-PSN and MA-DBN, respectively.
Fig.~6(b) depicts the sum rate versus the number of users when Data-SNR $=10$ dB. It is shown that the proposed MA-BFNet with MA-PSN and MA-DBN consistently outperforms the CEO and random selection with traditional WMMSE baselines, which demonstrates that the proposed framework can effectively scale to dense multiuser scenarios while maintaining a higher sum rate performance. 
To investigate the impact of CE accuracy on the beamforming performance, Fig.~6(c) compares the sum rate of the proposed MA-BFNet under different CE schemes when Pilot-SNR $=10$ dB. 
It is observed that using the CSI refined by MA-CENet yields the highest sum rate. The results confirm that the CE gain provided by MA-CENet can be effectively translated into a higher achievable sum rate when combined with MA-BFNet.

Furthermore, we evaluate the net sum rate versus the number of training positions $J$ used in the CE stage. We adopt a block-fading model where each coherence block consists of $T_{\text{total}} = 200$ normalized symbol intervals. Therefore, the net sum rate is defined as 
\begin{equation}
\begin{aligned}
\label{equ:s}
    R_{\text{net}} = \bigl(1 - \tfrac{J}{T_{\text{total}}}\bigr)\ R,
\end{aligned}
\end{equation}
where $R$ denotes the sum rate achieved by the corresponding beamforming scheme when the channel is estimated using $J$ positions. Fig.~6(d) plots the net sum rate versus $J$ when Data-SNR $= 10$ dB. It can be seen that the performance gain gradually saturates and then decreases as $J$ grows, revealing a fundamental trade-off between channel training overhead and data transmission efficiency in MA systems.

In addition, Table I compares the computational complexity of the proposed framework and the baseline networks in terms of FLOPs and runtime. Although MA-CENet and MA-BFNet exhibit higher complexity, the associated overhead is acceptable in practice when considering the approximately 3 dB NMSE improvement achieved within the all range of Pilot-SNR and the consistently superior sum-rate performance across the entire Data-SNR regime, as demonstrated in the preceding simulation results. 







\begin{table}[t]
\centering

\caption{{Neural Network Complexity Comparison}}
\label{tab:complexity-comparison}

\resizebox{\columnwidth}{!}{%
\begin{tabular}{lcccc}
\toprule
 & MA-CENet & MA-BFNet & DnCNN~\cite{ddn} & AdaFortiTran~\cite{ada} \\
\midrule
FLOPs (M)    & 93.72& 99.03& 63.62& 72.76\\
Runtime (ms) & 38.63& 22.82& 14.13& 14.54 \\
\bottomrule
\end{tabular}%
}
\end{table}


\section{Conclusion }


	This paper proposes a deep learning–based framework for multiuser multicarrier MA systems that jointly addresses CE, antenna positioning, and beamforming. First, MA-CENet enhances CSI reliability by integrating initial CE with Swin-Transformer denoising. Next, a Transformer encoder selects the optimal antenna position from candidate CSI sequences. Finally, a model-driven WMMSE algorithm designs beamforming to maximize the sum rate. Simulation results  demonstrate consistent performance gains over existing methods across various settings, particularly in wideband short-range multiuser scenarios requiring antenna mobility and high spectral efficiency. This work considers antenna movement at the BS and ignores antenna movement latency; future work will address the latency issue and consider the general setup with antenna movement at both the BS and users.

\vspace{15pt}
\color{red}
\end{document}